\documentclass[12pt]{article}

\pdfoutput = 1
\begin{document}
\title
{Constraints on the Generalized Uncertainty Principle from Black Hole Thermodynamics}

\author{
{\bf {\normalsize Sunandan Gangopadhyay}$^{a,b}
$\thanks{sunandan.gangopadhyay@gmail.com, sunandan@iucaa.ernet.in}},
{\bf {\normalsize Abhijit Dutta}$^{c}$\thanks{dutta.abhijit87@gmail.com}},
{\bf {\normalsize Mir Faizal}$^{d}$\thanks{f2mir@uwaterloo.ca}}\\
$^{a}${\normalsize Department of Physics, West Bengal State University, Kolkata 700126, India}\\$^{b}${\normalsize Visiting Associate, Inter
University Centre for Astronomy $\&$ Astrophysics, Pune, India}\\
$^{c}${\normalsize Department of Physics, Adamas Institute of Technology, Kolkata 700126, India}\\$^{d}${\normalsize Department of 
Physics and Astronomy, University of Waterloo, }\\
{\normalsize Waterloo, Ontario N2L 3G1, Canada}\\[0.3cm]
}
\date{}

\maketitle
\begin{abstract}
\noindent In this paper, we calculate the modification to the thermodynamics of a Schwarzschild black hole
in higher dimensions because of Generalized Uncertainty Principle (GUP).  
We use the fact that the leading order corrections to the entropy of a black hole has to be  
logarithmic in nature to restrict the form of GUP. We observe that in six dimensions, the usual GUP produces the correct 
form for the leading order corrections to the entropy of a black hole. However, in five and seven  dimensions a linear GUP, which is 
obtained by a combination of DSR with the usual GUP, is needed to produce the correct form of the corrections to the entropy of a 
black hole.  Finally, we demonstrate that in five dimensions, a  new form of  GUP containing quadratic and  cubic 
 powers of the momentum also produces the correct form for     the leading order corrections to the entropy of a black hole.
\end{abstract}
\section{Introduction}

The usual Heisenberg uncertainty principle is not consistent with the existence of a minimum measurable length. This is because 
according to the usual Heisenberg uncertainty principle
the  particle position can be measured to an arbitrary accuracy if the momentum 
is not measured. Hence, the idea of a position measurement within a minimum length scale, beyond which one cannot measure the 
position accurately, is absent within the usual Heisenberg uncertainty principle. 
The existence of a minimum length scale is present in all approaches to quantum gravity.  
The picture of spacetime as a smooth manifold breaks down at the Planck scale, 
and this in turn implies the restriction that at least the Planck length acts as a minimum measurable length scale. 
In fact, there are strong indications even from the physics of black holes, that 
any theory of quantum gravity should come equipped with a minimum length of the 
order of the Planck length \cite{z4,z5}. This is because the  energy needed to probe a region of spacetime below Planck length scale, 
will produce a mini black hole in that region of spacetime. This restricts the measurement of spacetime below the 
Planck length scale. The existence of a minimum length of the order of string length also occurs in string theory. 
As strings are the smallest probes that can be used to analyse any region of spacetime, it is not possible to probe 
spacetime below string length scale \cite{z2,zasaqsw,csdcas,cscds,2z}. It may be noted that a 
minimum length scale also exists in loop quantum gravity and leads to important phenomenological consequences.
In fact, this  minimum length in loop quantum gravity turns the big bang into a big bounce \cite{z1}.

It is possible to generalize the Heisenberg uncertainty principle to a Generalized uncertainty principle
(GUP) to make it consistent with the existence of a minimum  measurable length scale 
\cite{z2,zasaqsw,csdcas,cscds}. However, the deformation of the uncertainty principle leads to the deformation 
of the Heisenberg algebra, and this in turn deforms the coordinate representation of the momentum operators
\cite{2z,14,17,18,5,51,54,n7}. The deformation of the coordinate 
representation of the momentum operator produces correction terms for all quantum mechanical systems. 
Another deformation of the Heisenberg algebra is motivated by the study of a Doubly Special Relativity (DSR) \cite{2,21,3}.
In DSR, Planck energy is a universal constant like the velocity of light and is motivated by developments in 
discrete spacetime \cite{1q}, spontaneous symmetry breaking of Lorentz invariance in string
field theory \cite{2q}, ghost condensation \cite{q2},  spacetime foam models \cite{3q}, spin-network in loop quantum gravity \cite{4q},
noncommutative geometry \cite{5q}, 
and Horava-Lifshitz gravity \cite{6q}. 
It may be noted that DSR has been generalized to curved spacetime and the resultant theory  
is called gravity's rainbow \cite{n1,n2}. It has been possible to combine the deformation of the Heisenberg algebra coming from 
DSR with the deformation of the Heisenberg algebra coming from GUP \cite{main2}.
The resultant commutator between the coordinates and momentum operators 
contain  linear terms in the momentum, and this gives rise to non-local fractional derivative terms in all the 
quantum mechanical Hamiltonians except in one dimensional case. However, various
interesting one dimensional quantum mechanical systems have been studied using this deformed Heisenberg algebra
because such non-local terms do not exist for one dimensional Hamiltonians. The 
transition rate of ultra cold neutrons in gravitational field has been analysed using this 
deformed algebra \cite{n6}. It has been argued that the higher dimensional non-local Hamiltonians can be analysed 
using the theory of harmonic extension of functions \cite{mf, mf22}.  
 
The deformation of the Heisenberg uncertainty principle also modifies the thermodynamics of black holes 
This modification has been  calculated for the Schwarzschild and Reissner-Nordstr\"{o}m black hole in four dimensions  
\cite{ca, ca00, ca22, ca66, ca44, sgad}. This was done by first writing expressing the 
bound on the maximum momentum in GUP, in terms of a bound on  maximum energy of the system. 
This energy was then related to the  energy of an emitted photon, and thus it could be related to the 
temperature of the black hole. The uncertainty in the position was then taken to be 
proportional to the radius of the event horizon of the black hole. This led to the 
modification to the temperature of the black hole from GUP. 

The existence of a logarithmic correction to the entropy of a black hole is a universal predication coming from all 
approaches which analyze leading order corrections to the entropy of a black hole \cite{black, black1, black4, black6, black7}. 
  In fact, it has been argued that logarithmic correction to the entropy would be the leading order 
  correction to the entropy of any thermodynamic system due to small statistical fluctuations around equilibrium, and 
  as black holes are thermodynamic systems, the leading order corrections to the entropy of a black hole has to have the 
  form of a logarithmic correction \cite{77}. It may be noted that there is an ambiguity regarding the coefficient to this 
  logarithmic correction to the entropy of a black hole, and this coefficient depends on the details of the model being considered. 
  In this paper, we will study the thermodynamics of black holes in higher dimensions using the GUP. 
  The important ingredient in our analysis is to use the fact that the leading order 
  correction to the entropy has to have a logarithmic form, which in turn fixes the form of the GUP. 
  This essentially amounts to reversing the argument used in previous studies. 
  It is observed that in six dimensions, the usual GUP leads to the correct form for the leading order correction 
  to the entropy of a black hole. However, in five and seven 
  dimensions, one requires a linear GUP (which is obtained by a combination of the regular 
  GUP with DSR) to produce the correct form for the leading order corrections 
  to the entropy of a black hole. 
  We also propose a new form of    GUP which contains cubic and quadratic   powers of the momentum. The correction to the entropy of a black hole in 
  five dimensions corresponding to this new GUP is also obtained.

The paper is organized as follows. In section 2, we present a discussion of the Schwarzschild black hole
in higher dimensions, and obtain an expression for black remnants and critical masses in various dimensions.
In section 3, we restrict   the form
of the GUP by constraining it to produce the 
correct form of the correction to the entropy of a black hole. Finally, we  conclude this paper in section 4.   

\section{Schwarzschild black hole in higher dimensions}
In this section, we will analyze the corrections to the thermodynamics of Schwarzschild black hole in $d$-dimensions
using the simplest form of the GUP. The relation between the
mass and temperature of a black hole with the effect of the GUP taken into account will be used for calculating the 
heat capacity of the black hole in higher dimensions. We will then use this result to 
calculate the entropy of the black hole. In the subsequent discussion, we will use 
natural units $c=1=\hbar$ and $k_B=1$.

We start our analysis by considering a  a Schwarzschild black hole in $d$-dimensions with mass $M$. 
The metric for this black hole can be written as \cite{emparan, aman}
\begin{eqnarray}
ds^2 = -\left(1-\frac{\mu}{r^{d-3}}\right)dt^2+\frac{1}{\left(1-\frac{\mu}{r^{d-3}}\right)}dr^2+r^2d\Omega^2_{d-2}
\label{metric}
\end{eqnarray}
where $\mu$ is related to mass as follows 
\begin{eqnarray}
\mu=\frac{16\pi G_dM}{(d-2)\Omega_{d-2}}~. 
\label{mu}
\end{eqnarray}
Here we have used the Newton's constant $G_d$ in $d$-dimensions, which is related to the Planck mass as
\begin{eqnarray}
G_d=\frac{1}{M_p^{d-2}}~. 
\label{G_d}
\end{eqnarray} 
We also need the   volume of $(d-2)$ unit sphere, which is given by 
\begin{eqnarray}
\Omega_{d-2}=\frac{2\pi^{(\frac{d-1}{2})}}{\Gamma(\frac{d-1}{2})}~.
\label{omega}
\end{eqnarray}
The horizon radius $r_h$ can be calculated by solving the equation  
\begin{eqnarray}
1-\frac{\mu}{r_h^{d-3}}=0. 
\end{eqnarray}
Thus, we obtain 
\begin{eqnarray}
r_h=\mu^{\frac{1}{d-3}}=\left[\frac{16\pi M}{(d-2)M_p^{d-2} \Omega_{d-2}}\right]^{\frac{1}{d-3}}.
\label{horizon radius}
\end{eqnarray}
Near the horizon the  uncertainty in the position of an emitted particle will be order of the Schwarzschild radius. 
So we can write 
\begin{eqnarray}
\delta x=\epsilon r_h
\label{pos_uncrnty}
\end{eqnarray}
where $\epsilon$ is a calibration factor.
Furthermore, uncertainty in the momentum for the particle can be written as
\begin{eqnarray}
\delta p = T.
\label{mom_uncrnty}
\end{eqnarray}
As we are carrying out our calculations in the thermodynamic equilibrium, the temperature of the emitted particle 
will be identical to the temperature of the black hole.

The usual GUP is given by \cite{adler}
\begin{eqnarray}
\delta x\delta p\geq\frac{\hbar}{2}\left\{1+ \frac{\beta^2 l_p^2}{\hbar^2}(\delta p)^2\right\}
\label{gup}
\end{eqnarray}
where $l_p$ is the Planck length ($\sim 10^{-35}m$) and $\beta$ is a dimensionless constant.
Therefore, we can write 
\begin{eqnarray}
\epsilon \left[\frac{16\pi M}{(d-2)M_p^{d-2} \Omega_{d-2}}\right]^{\frac{1}{d-3}} T = \frac{1}{2}\left\{1+{\beta^2 l_p^2}T^2\right\}.
\label{gup_mod}
\end{eqnarray}
Neglecting the effects coming from the GUP, we could write this expression as 
\begin{eqnarray}
\epsilon \left[\frac{16\pi M}{(d-2)M_p^{d-2} \Omega_{d-2}}\right]^{\frac{1}{d-3}} T = \frac{1}{2}~.
\label{gup_no_grav}
\end{eqnarray}
The temperature  of the black hole is given by \cite{Ang}
\begin{eqnarray}
T = \frac{1}{4\pi}\sqrt{A_{,r}(r_h)B_{,r}(r_h)}
\label{tp1}
\end{eqnarray}
where $A(r)$ and $B(r)$ are the coefficients appearing in the metric of the black hole   
\begin{eqnarray}
ds^2 = -A(r)dt^2 + \frac{1}{B(r)}dr^2 + h_{ij}dx^idx^j.
\label{mr1} 
\end{eqnarray}
For Schwarzschild black hole, we have  
\begin{eqnarray}
A(r) = B(r) = 1-\frac{\mu}{r^{d-3}}
\label{tp2}
\end{eqnarray}
and hence the temperature  for the  Schwarzschild black hole is given by 
\begin{eqnarray}
T = \frac{d-3}{4\pi}\frac{1}{\mu^{\frac{1}{d-3}}}~. 
\label{tm1}
\end{eqnarray}
Now using this expression for the temperature of the Schwarzschild black hole in Eq. (\ref{gup_no_grav}), we obtain  
\begin{equation}
\epsilon = \frac{2\pi}{d-3}~. 
\end{equation}
Substituting the value of $\epsilon$ in Eq. (\ref{gup_mod}), yields the relation between mass and temperature of the black hole 
\begin{eqnarray}
M = a'\left[\frac{1}{T}+\frac{\beta^2}{M_p^2}T\right]^{d-3}
\label{mass-temp}
\end{eqnarray}
where 
\begin{equation}
 a' = \left(\frac{d-3}{4\pi}\right)^{(d-3)}\left[\frac{(d-2)\Omega_{d-2}M_p^{d-2}}{16\pi}\right]
\end{equation}
and we have used $l_p = {M_p}^{-1}$~.

Now we can calculate corrections to the heat capacity of the Schwarzschild black hole from the usual GUP as follows 
\begin{eqnarray}
C = \frac{dM}{dT} = a'(d-3)\left[\frac{1}{T}+\frac{\beta^2}{M_p^2}T\right]^{d-4}\left[-\frac{1}{T^2}+\frac{\beta^2}{M_p^2}\right].
\label{heat capacity}
\end{eqnarray}
The remnant mass (where the evaporation stops) for the black hole can be obtained by setting $C = 0$ :  
\begin{eqnarray}
M_{rem} = a' \left(\frac{2\beta}{M_p}\right)^{(d-3)}.
\label{remnant mass}
\end{eqnarray}
Furthermore, the temperature can be expressed in terms of mass as
\begin{eqnarray}
T = \frac{M'M_p^2 \pm \sqrt{M_p^4 M'^2  - 4\beta^2 M_p^2}}{2\beta^2}
\label{hcp}
\end{eqnarray}
where 
\begin{eqnarray}
 M' = \left(\frac{M}{a'}\right)^{(3-d)}
\end{eqnarray}
The temperature 
has to be a real quantity, so  the critical mass is given by 
\begin{eqnarray}
M_{cr} = a' \left(\frac{2 \beta}{M_p}\right)^{(d-3)}.
\label{critical mass}
\end{eqnarray}
Thus, we have demonstrated that the remnant mass of the black hole is equal to its critical mass.


\section{Black hole thermodynamics in various dimensions }
With the general formalism in place, we will now study the thermodynamics of black holes in different dimensions. 
The universality of logarithmic corrections as the leading order corrections to the entropy of a black hole will be used
to fix the form of the GUP in various dimensions. 
We will first calculate the corrections to the thermodynamics in five dimensions. 
The heat capacity in five dimensions can be explicitly written as 
\begin{eqnarray}
C = 2a'\left(-\frac{1}{T^3}\ + \frac{\beta^4}{M_p^4}T\right).
\label{cp1}
\end{eqnarray} 
The entropy can be calculated using the heat capacity of this black hole as follows 
\begin{eqnarray}
S = \int{C\frac{dT}{T}}~.
\label{en1}
\end{eqnarray}
In five dimensions this yields 
\begin{eqnarray}
S = 2a'\left(\frac{1}{3T^3}\ + \frac{\beta^4}{M_p^4}T\right). 
\label{en2}
\end{eqnarray}
However, this cannot be an acceptable form for the correction of the black holes entropy because it is known that the leading 
order correction to   entropy of a black hole is a logarithmic correction. Thus, we need to consider a more general form of GUP 
to obtain the correct form of the entropy of a black hole in five dimensions. So, we use the
  linear GUP, which is 
 obtained by combining DSR with the usual GUP, for this purpose    \cite{main2} 
\begin{eqnarray}
\delta x\delta p&\geq&\frac{\hbar}{2}\left\{1-\frac{\alpha l_p}{\hbar}\delta p +\frac{\beta^2 l_{p}^2}{\hbar^2}(\delta p)^2\right\}.
\label{gup-1}
\end{eqnarray} 
Now this linear GUP modifies the  relation between the  mass and temperature of a black hole 
\begin{eqnarray}
\sqrt{\mu} = \frac{1}{2\pi}\left(\frac{1}{T}-\frac{\alpha}{M_p}+\frac{\beta^2}{M_p^2}T\right).
\label{mt1}
\end{eqnarray}
So, the  heat capacity can be written as 
\begin{eqnarray}
C = 2a'\left(-\frac{1}{T^3}+\frac{\alpha}{M_p^4}\frac{1}{T^2}-\frac{\alpha \beta^2}{M_p^3}+\frac{\beta^4}{M_p^4}T\right). 
\label{mt2}
\end{eqnarray}
Using Eq. (\ref{en1}), we can obtain an expression for the entropy  of the black hole 
\begin{eqnarray}
S = 2a'\left(\frac{1}{3T^3}-\frac{\alpha}{2M_p}\frac{1}{T^2}-\frac{\alpha \beta^2}{M_p^3} \ln T +\frac{\beta^4}{M_p^4}T\right). 
\label{mt3}
\end{eqnarray}
Finally, using the relation between mass and temperature given by Eq. (\ref{mt1}), 
we can recast the entropy of a five dimensional Schwarzschild black hole as 
\begin{eqnarray}
S &=& \frac{A}{4{l_p}^3}+\frac{3\alpha}{(128\pi)^{\frac{1}{3}}}
\left(\frac{A}{4{l_p}^3}\right)^{\frac{2}{3}}
-\frac{3\beta^2}{(256\pi^2)^{\frac{1}{3}}}\left(\frac{A}{4{l_p}^3}\right)^{\frac{1}{3}}
\nonumber \\ && -\frac{3\alpha\beta^2}{16\pi}\ln\left[\frac{\alpha}{M_p}+\frac{(16\pi)^{\frac{1}{3}}}{M_p}\left(\frac{A}{4{l_p}^3}\right)^{\frac{1}{3}}\right]
\label{mt4}
\end{eqnarray}
where  the area is expressed as  $A = \Omega_{d-2}{r_h}^{(d-2)}$.

Now we will analyse the results in six dimensions. The expression for the heat capacity in six dimensions from the usual GUP 
can be obtained from Eq. (\ref{heat capacity}) is given by 
\begin{eqnarray}
C = 3a'\left(-\frac{1}{T^4}-\frac{\beta^2}{M_p^2}\frac{1}{T^2}+\frac{\beta^4}{M_p^4}\right).
\label{zc1}
\end{eqnarray}
Hence, the corrections to the entropy of the six dimensional Schwarzschild black hole from the usual GUP reads 
\begin{eqnarray}
S = \frac{A}{4{l_p}^4}-\beta^2\frac{3\sqrt{3}}{4\sqrt{2}\pi}\left(\frac{A}{4{l_p}^4}\right)^{\frac{1}{2}}-\frac{\beta^4}{M_p^4}-
\frac{\beta^4}{M_p^4}\ln\left[\left(\frac{128\pi^2}{M_p}\right)^{\frac{1}{4}}l_p\left(\frac{A}{4{l_p}^4}\right)^{\frac{1}{4}}\right].
\label{yen}
\end{eqnarray}
Even though this is an acceptable form for the entropy of a black hole, we will calculate the corrections to the entropy of
a six dimensional Schwarzschild black hole from the linear GUP for completeness. 
The corrections to the entropy of the six dimensional Schwarzschild black hole from the linear GUP reads 
\begin{eqnarray}
S=\frac{A}{4l_p^4}+\frac{\alpha}{\sqrt{\pi}}\left(\frac{2}{3}\right)^{\frac{1}{4}}\left(\frac{A}{4{l_p}^4}\right)^{\frac{3}{4}}
-\frac{(\alpha^2-\beta^2)}{\pi}\frac{3\sqrt{3}}{4\sqrt{2}}\left(\frac{A}{4{l_p}^4}\right)^{\frac{1}{2}}-\frac{27\alpha^2\beta^2}{16 \pi^2}
\nonumber\\
+\frac{27\alpha^2\beta^2}{32 \pi^2}\ln\left[\frac{\alpha}{M_p}+\frac{4 \pi}{3M_p}\left(\frac{3}{2}\right)^{\frac{1}{4}}
\left(\frac{A}{4{l_p}^4}\right)^{\frac{1}{4}}\right]. 
\label{ye1}
\end{eqnarray}
Finally, we study a seven dimensional black hole. Following a similar procedure, we obtain the corrections to the entropy of a black hole 
from the usual GUP in seven dimensions to be  
\begin{eqnarray}
S=\frac{A}{4l_p^5}-\frac{5\beta^2}{(4\pi)^{\frac{4}{5}}}\left(\frac{A}{4{l_p}^5}\right)^{\frac{3}{5}}+5\beta^2\frac{M_p^3}{6\pi^2}~.
\label{ye2}
\end{eqnarray}
Once again this is not an acceptable form for the corrections to the entropy of a black hole 
as the leading order corrections have to be logarithmic in nature. 
So, we again consider a linear GUP given by Eq. (\ref{gup-1}), and calculate the  
corrections to the entropy of a seven dimensional  Schwarzschild black hole from this linear  GUP 
\begin{eqnarray}
S&=& \frac{A}{4l_p^5}-5\alpha\frac{(2)^{\frac{3}{5}}}{(\pi)^{\frac{2}{5}}}\left(\frac{A}{4{l_p}^5}\right)^{\frac{4}{5}}
+\frac{5\beta^2}{3(2\pi)^{\frac{4}{5}}}\left(\frac{A}{4{l_p}^5}\right)^{\frac{3}{5}}
\nonumber \\ && +\frac{45\alpha\beta^2}{(2)^{\frac{11}{5}}(\pi)^{\frac{6}{5}}}\left(\frac{A}{4{l_p}^5}\right)^{\frac{2}{5}}
+\frac{5\beta^4}{(2)^{\frac{8}{5}}(\pi)^{\frac{3}{5}}}\left(\frac{A}{4{l_p}^5}\right)^{\frac{1}{5}}\nonumber\\
&&-\frac{19\alpha\beta^4}{8 \pi^2}
+\frac{3\alpha\beta^4}{4 \pi^2}\ln\left[\frac{\alpha}{M_p}+(2\pi)^{\frac{2}{5}}\left(\frac{A}{4{l_p}^5}\right)^{\frac{1}{4}}\right]. 
\label{ven1}       
\end{eqnarray}
This is an acceptable form for the corrections to the entropy of a black hole. 
Thus, we conclude that in five and seven dimensions, we need a linear GUP to obtain the correct form for the corrections 
to the entropy of a black hole. It should also be noted that both the linear and quadratic correction terms in the 
momentum uncertainty are required to get the logarithmic correction. However, in even dimensions, the usual GUP
is enough to generate the correct form for the corrections to the entropy of a black hole. 

Finally,   we make another interesting observation. In absence of a linear term in GUP in five dimensions,
the correct form for the leading order 
corrections to the entropy of a black hole can also be generated from a cubic power of momentum in the GUP. 
Thus, we propose a modified GUP which contains a cubic and quadratic powers of the momentum, 
\begin{eqnarray}
\delta x\delta p&\geq&\frac{\hbar}{2}\left\{1+\frac{\beta^2 l_{p}^2}{\hbar^2}(\delta p)^2+\frac{\gamma^3 l_{p}^3}{\hbar^3}(\delta p)^3\right\}
\label{GUP_higher order 1}
\end{eqnarray}
where $\gamma$ is a suitable parameter in the theory. 
The corrections to the entropy corresponding to this cubic form of GUP can now be written as 
\begin{eqnarray}
S &=& \frac{A}{4{l_p}^3}-\frac{3\beta^2}{(16\pi)^{\frac{2}{3}}}\left(\frac{A}{4{l_p}^3}\right)^{\frac{1}{3}}+
\frac{33\beta^2\gamma^3}{2(16\pi)^{\frac{5}{3}}}\left(\frac{A}{4{l_p}^3}\right)^{-\frac{2}{3}} \nonumber \\ && 
 -\frac{3\gamma^3}{16\pi}-\frac{2\gamma^3}{16\pi}\ln(4\pi)
-\frac{2\gamma^3}{16\pi}\ln A.
\label{qen}
\end{eqnarray}
It may be noted that the coefficient of the logarithmic term depends upon the coefficient of the cubic power of the  momentum  in
this new GUP. 
Thus,  in five dimensions a  logarithmic correction to the entropy  
can be obtained from an cubic or a linear power of momentum  in the GUP.

\section{Conclusions}
In this paper, we calculate the leading order corrections to the entropy of a black hole in higher dimensions 
using a general form for the GUP. We use the fact that the leading order corrections to the entropy of any thermodynamic system are 
logarithmic in nature to constraint the form of the GUP that has to be used. 
We observe that in six dimensions the usual GUP produces the correct leading order correction term for 
the black hole entropy. However, in the five and seven  dimensions we have to use the linear GUP, which is obtained by 
combining the regular GUP with DSR, to obtain the correct form of the leading order correction to the black hole 
entropy. 
Furthermore, it is known that in four dimensions the usual GUP produces the correct form  
for the corrections to the entropy of a black hole \cite{ca44}. We also proposed a new form of GUP which contains cubic and 
quadratic powers of momentum. It was also demonstrated that this new form of GUP also produces  the correct 
form for the correction  of the entropy of a black hole in five dimensions. 
This suggests that in all even dimensions the usual form 
of GUP might be  enough to produce the correct form for the the corrections to the entropy of a black hole. However, in odd dimensions
an odd power of momentum in the GUP might be needed to produce 
the correct form for the corrections to the entropy of a black hole.  It will be interesting to investigate this further in various dimensions, 
and with various modifications to the GUP. It will also be interesting to try to find a general argument for this to be the case 
in all odd and even dimensions.

It may be noted that there is an arbitrariness in the exact form of GUP, and there is a large class of 
phenomenologically valid form that produce the correct physical effects. Thus, we need a selection produce to reduce this large 
number of phenomenologically valid forms of GUP. Black hole thermodynamics seems to be that  selection 
procedure. This is the first time that the form of GUP in different dimensions has been argued using 
black hole thermodynamics. The previous works in this direction, ended up calculating the corrections to the black hole 
thermodynamics using a selected form of GUP. 
It may also be noted that in higher dimensions other interesting solutions like black rings and black Saturn can exist. 
Corrections to such systems in higher dimensions using gravity's rainbow has been recently studied \cite{gr, rg}. 
It would be interesting to repeat the present analysis for these objects, and analyse if the form of GUP fixed by a 
Schwarzschild black hole produces the correct form of the leading order corrections to the  
entropy of these black objects. Finally, we have also proposed a  new GUP, which contains cubic and quadratic 
powers of momentum. It will be interesting to calculate the Heisenberg algebra corresponding to this new GUP, 
and the coordinate representation of the momentum operator corresponding to this new GUP. This new GUP will also deform 
all the quantum mechanical Hamiltonians. It will be interesting to investigate the phenomenological consequence of this 
GUP.  In fact, it is expected that this new GUP with cubic and quadratic powers of momentum might lead to a discretization  
of space and time, as the linear GUP has lead to a discretization of space \cite{l1, l2}, and a modification of the Wheeler-DeWitt equation motivated 
by linear GUP has lead to a discretization in time \cite{f}.

\end{document}